\documentclass[11pt,a4paper]{article}
\usepackage{geometry}
\usepackage[round]{natbib}
\usepackage{authblk}
\usepackage{graphicx}
\usepackage{tabularx}
\usepackage{multirow}
\usepackage{multicol}
\usepackage{amsthm}
\usepackage{algorithm2e}
\usepackage{ulem}
\usepackage{color}
\usepackage{amsmath}
\usepackage{cite}
\usepackage[caption=false]{subfig}

  \usepackage{lipsum}

\usepackage{ragged2e}
\usepackage{sectsty}
\sectionfont{\fontsize{15}{15}\selectfont} 
\subsectionfont{\fontsize{13}{15}\selectfont}
\usepackage{tikz} 
\usetikzlibrary{shapes,arrows}

\usepackage{setspace}
\usepackage{hyperref}

 \newgeometry{a4paper, total={170mm,257mm},left=25mm,right=25mm,bottom=25mm,top=25mm,}

\begin{document}
\pagenumbering{gobble}
\title{\LARGE \textbf{Evaluating bird collision risk of a high-speed railway crossing the habitat of the crested ibis (\textit{Nipponia nippon}) in Qinling Mountains, China}}
\centering
\author[1]{\large Han Hu }
\author[2]{Junqing Tang\thanks{Co-first author: junqing.tang@frs.ethz.ch (have equivalent contribution as the first author)}}
\author[2]{Yi Wang}
\author[1]{Hongfeng Zhang}
\author[1]{Dong Wu}
\author[3]{Yingchun Lin}
\author[1]{Lina Su}
\author[1]{Yan Liu}
\author[4]{Wei Zhang}
\author[5]{Chao Wang}
\author[1]{Xiaomin Wu\thanks{Corresponding author: xiaominwu66@126.com}}

\affil[1]{ \textit{\small Shaanxi Institute of Zoology, Xi'an, 710001, China}}
\affil[2]{ \textit{\small ETH Zurich, Future Resilient Systems, Singapore-ETH Centre, 1 CREATE Way, CREATE Tower, 138602 Singapore}}
\affil[3]{ \textit{\small Fourth Research Institute of Telecommunications Technology Co., Ltd., Xi'an, 710061, China}}
\affil[4]{ \textit{\small Xi'an Branch of Chinese Academy of Sciences, Shaanxi Academy of Sciences, Xi'an, 710061, China}}
\affil[5]{ \textit{\small Administration of Hanzhong National Nature Reserve for Crested Ibis, Hanzhong, 723399, China }}

\renewcommand\Authands{ }
\renewcommand\Authsep{  }
\renewcommand\Authfont{\bfseries}

\date{}

\maketitle

\begin{abstract}
Bird collisions with high-speed transport modes is a vital topic on vehicle safety and wildlife protection, especially when high-speed trains, with an average speed of 250km/h, have to run across the habitat of an endangered bird species. This paper evaluates the bird-train collision risk associated with a recent high-speed railway project in Qinling Mountains, China, for the crested ibis (\textit{Nipponia nippon}) and other local bird species. Using line transect surveys and walking monitoring techniques, we surveyed the population abundance, spatial-temporal distributions, and bridge-crossing behaviors of the birds in the study area. The results show that: (1) The crested ibis and the egret were the two most abundant waterfowl species in the study area. The RAI of these two species were about 43.69\% and 42.91\%, respectively; (2) Crested ibises overall habitat closer to the railway bridge. 91.63\% of them were firstly detected within the range of 0m to 25m of the vicinity of the bridge; (3) the ratio between crossing over and under the railway bridge was about 7:3. Crested ibises were found to prefer to fly over the railway bridge (89.29\% of the total crossing activities observed for this species). Egrets were more likely to cross the railway below the bridge, and they accounted for 60.27\% of the total observations of crossing under the bridge. We recommend that, while the collision risk of crested ibises could be low, barrier-like structures, such as fences,  should still be considered to promote the conservation of multiple bird species in the area. This paper provides a practical case for railway ecology studies in China. To our best knowledge, this is the first high-speed railway project that takes protecting crested ibises as one of the top priorities, and exemplifies the recent nationwide initiative towards the construction of ``eco-civilization" in the country.

\vspace{0.5cm}
\raggedright
\text{\textit{Keywords:}}
Ecological civilization; Crested ibis; Railroad; Bird collision; High-speed railway.
\end{abstract}


\setlength\parindent{2em}\justify

 \newgeometry{a4paper, total={170mm,257mm},left=40mm,right=40mm,bottom=35mm,top=35mm,}
\pagenumbering{arabic}
\setlength\parindent{2em}\justify 

\doublespacing

\section{Introduction}
Hundreds of millions of wild birds are killed each year due to collisions with different types of man-made structures, such as buildings, power lines, wind turbines, and vehicles~\citep{loss2014bird,erickson2005summary,manville2005bird,desholm2005avian,huppop2006bird}. Among them, vehicle collisions are one of the top sources. Recent quantitative reviews on North American and European bird-vehicle collision literature estimated that between 89 and 340 million birds are killed annually from vehicle collisions on U.S. roads~\citep{loss2014estimation}, and the estimated mortality on European roads was about 0.35 to 27 million per year~\citep{erritzoe2003bird}. Railways collisions are also documented to account for a large portion of anthropogenic mortality of birds~\citep{popp2017railway}. The high-speed railway is of great interest to many countries such as China. According to a new national railway network plan issued by Chinese government in 2016~\citep{NRA}, the total mileage of high-speed railway lines will achieve 30,000 km by 2020 and 45,000 km by 2030, connecting all the capital cities (provincial capitals, excluding Lhasa, the capital of Tibet Autonomous Region) and cities of a population of over 0.5 million~\citep{zhou2018implications,wang2018potential}. With such fast-expanding railway infrastructure, the conservation on ecology and wildlife have become increasingly prominent in China~\citep{wang2015china}. However, railway ecology has been underdeveloped and not many studies have contributed to analyzing bird collisions associated with railways~\citep{loss2015direct,popp2017railway,godinho2017bird}.


The Xi'an-Chengdu High-speed Railway (XCHR) is a recent project that connects the capital of Shaanxi Province, Xi'an, and the capital of Sichuan Province, Chengdu, with an average design speed of 250 km/h. As one of the most important projects under China's Belt and Road initiatives~\citep{BeltRoad}, it provides a vital connector for northwest and southwest of the country. Moreover, it is the first north-south high-speed railway that crosses the Qinling Mountains (an east-west mountain range that lies in between of these two provinces). This mountain range is known to be the main habitat of several endangered and rare species, including takin (\textit{Budorcas taxicolor}), giant panda (\textit{Ailuropoda melanoleuca}), crested ibis (\textit{Nipponia nippon}), and golden snub-nosed monkey (\textit{Rhinopithecus roxellana})~\citep{Panda}. This to-be-built XCHR runs across a National Nature Reserve established for the crested ibis, wherein the birds are expected to cross the railway area on a daily basis~\citep{dias2006distance,godinho2017bird}. It was feared that the avians in the reserve, especially this endangered bird species, could be exposed to a high risk of collisions with circulating bullet trains after the completion of the XCHR. 

The crested ibis is an endangered bird species that used to be widespread in China, Japan, Korea, and Russia~\citep{feng2019genomic,li2014genomic}. Continuous habitat loss, population isolation, and human interference have once brought this species to the brink of extinction. In 1981, the last seven wild crested ibises were rediscovered at Yangxian county in Shaanxi province~\citep{liu1981rediscovery,xiao2009return}. China has invested a significant amount of efforts into the conservation of this endangered bird species ever since. As a result, its wild population had raised to about 40 in the late 1990s~\citep{li1998current}, and the total population now has reached to more than 2000~\citep{feng2019genomic}. Various protection measures have been applied to the crested ibis, including extensive captive breeding and reintroduction in other suitable habitats in China and Japan~\citep{chen2013phylogenetic}. To date, they are listed at the ``Top" protection level in the Wildlife Protection Law of the People's Republic of China~\citep{wang2015china,kong2013road} and also at the ``Endangered" level on the Red List established by International Union for Conservation of Nature (IUCN)~\citep{IUCN}.

In the context of bird collision studies, the majority have been dedicated to investigate bird mortality associated with traffic on roads~\citep{santos2016avian,husby2017traffic}. The state-of-art knowledge about railway ecology associated with birds is scant, and the relevant studies are much less developed with respect to high-speed railways because of their limited and relatively recent presence~\citep{dorsey2015ecological}. In fact, railway ecology is one of the emerging disciplines~\citep{barrientos2019railway} and has been relatively underdeveloped in the study of transportation ecology~\citep{popp2017railway}. One significant contribution to this field is the book "Railway Ecology" published in 2017~\citep{raiwayecology}, which provided a unique overview of the impacts of railways on biodiversity. In the book,~\citet{godinho2017bird} carried out a study of the mortality and bridge-crossing behaviors of birds near a railway crossing a wetland landscape in Portugal. The target species were passerines and waterfowls in the area and they found that the passerines were more vulnerable in this case. In addition to this book, two studies from Spain also contributed to the study of bird collisions with high-speed railways.~\citet{garcia2017board} studied the avian mortality caused by a Spanish high-speed railway using on-board camera recordings. The bird-train collision mortality was estimated as 60.5 birds/km/year on a line section of 53 trains/day and 26.1 birds/km/year in a section of 25 trains/day. However, the bird species examined were mainly passerines and raptors.~\citet{malo2017cross} conducted an investigation on bird behaviors around a 22-km high-speed railway in Spain from 2011 to 2015. It revealed how bird species responded to high-speed trains at various scales, and showed how the infrastructure impacts bird communities due to both the habitat changes and the increases in mortality risk. In addition, studies on bird mortality in railway ecology from China have been even more deficient and less reported in the literature~\citep{peng2007building,piao2016preliminary}. \color{black}

This paper contributes to filling the following two research gaps: (1) Studies of railway ecology on the crested ibis is relatively less developed. Also, (2) Evaluation of bird-train collision risk associated with high-speed railways in China need to be addressed for future research and implications. Therefore, we conduct this study with the focus on evaluating bird-train collision risks for the crested ibises and other local bird species before the operation of the XCHR service. This paper demonstrates China's commitment in the protection of endangered wildlife species in its expanding infrastructure projects. To our best knowledge, this is the first railway ecology study associated with the crested ibis.

\section{Study area}
Railway projects are accomplished by the following stages in a project development cycle: inception, planning, design, construction, operation, and maintenance~\citep{roberts2015incorporating}. Our investigation was carried out during the late construction stage. After the completion of the main engineering structures of the XCHR (roadbeds, tunnels, bridges, etc.), our work took place from March to July 2016, before the completion of the track laying construction.

\begin{figure}[h!]
\centering
\includegraphics[width=1\textwidth]{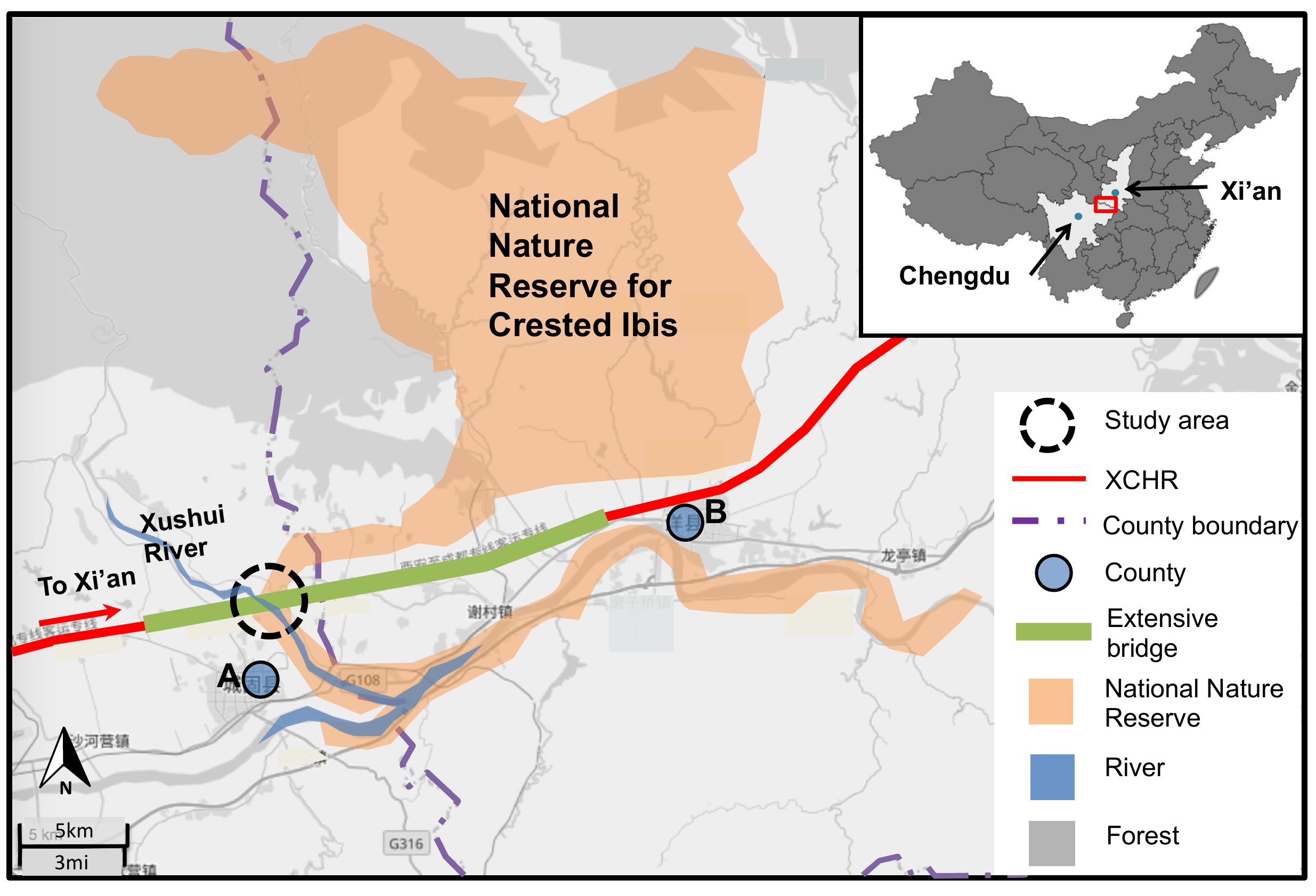}
\caption{\small Geographical layout of the study area (Source: OpenStreetMap\textcopyright).  Note that the B county is the famous Yangxian county. (color is needed)}
\label{fig_1}       
\end{figure}

As shown in Fig.~\ref{fig_1}, the national nature reserve for crested ibis is surrounding the County B in Shaanxi province. The reserve area is occupied by forest and wetland, such as rivers, paddy fields, and small ponds. For safety reasons, the alignment of the high-speed railway has to keep reasonably straight and smooth. This constraint has led XCHR to unavoidably cross the nature reserve with a length of 1.6 km in the section over the Xushui river, which is the target study area in this study (highlighted in the black circle in Fig.~\ref{fig_1}). The crossing is made through an extensive bridge with a length of 27km (the green strip in Fig.~\ref{fig_1}), and an average elevated height of 10 m.

According to the local wildlife censuses, there are 37 families of 205 bird species documented in this area. In addition to the crested ibis, most of the bird species are waterfowls and passerines, such as egret (\textit{Ardea alba}), grey heron (\textit{Ardea cinerea}), white wagtails (\textit{Motacilla alba}), and swallows (\textit{Hirundo rustica}). Table~\ref{tab3} summarizes the main species of medium-sized and large birds in the area and their protection classes according to the new Wildlife Protection Law of People's
Republic of China and the IUCN's Red List. Except the crested ibis, other species are all in the ``Least concern" class and not particularly protected by China's national law.

\begin{table}[hpt!]
\caption{\small Main medium-sized and large birds in the study area and their protection levels according to China's national law and the IUCN's Red List}
\label{tab3}
\resizebox{\textwidth}{!}{\begin{tabular}{|l|l|l|l|}
\hline
\centering
 \textbf{Scientific name} & \textbf{Common name} & \textbf{Protective level} & \textbf{IUCN} \\ \hline
  Nipponia nippon & crested ibis, Asian crested ibis & I & EN \\ \hline 
  Ardea alba & egret, white egret & - & LC \\ \hline 
  Ardea cinerea & grey heron & - & LC \\ \hline 
  Pica pica & common magpie, Eurasian magpie & - & LC \\ \hline
  Nycticorax nycticorax & night heron, black-crowned night heron & - & LC \\ \hline
  Streptopelia decaocto & Eurasian collared dove & - & LC \\ \hline 
  Ardeola bacchus & Chinese pond heron & - & LC \\ \hline
  Phalacrocorax carbo & cormorant, great cormorant & - & LC \\ \hline 
  Bubulcus ibis & cattle egret & - & LC \\ \hline
\end{tabular}}
* \small{Ex-Extinct; EW-Extinct in the wild; CR-Critically endangered; EN-Endangered; VU-Vulnerable; NT-Near threatened; LC-Least concern.}
\end{table}

\section{Methodology}
In this study, we used transect surveys and walking monitoring techniques to estimate the population abundance, spatial-temporal distributions, and bridge-crossing behaviors of the birds in the study area. The abundance of animal populations can be estimated by applying the \textit{line transect method}, or line-intercept sampling technique~\citep{kaiser1983unbiased}. In general, one or more transect lines of fixed length are established in the area. The line is walked, at least once, and the data on birds observed are recorded. In bird studies, it is common to establish a bandwidth distance on either side of the line and only record birds observed within this distance. We planned two 2 km line transects immediately next to both sides of the railway bridge (in parallel) with a bandwidth of 100m, as suggested by~\citet{malo2017cross}, to cover a sufficient vicinity of the bridge. In the end of the survey, we collected the frequency of the observations of encountered species for their corresponding abundance estimation~\citep{bibby2000bird}.

Observers used telescopes to conduct field monitoring simultaneously in two groups, with at least three observers in each. They rotated between those two transects every one hour during each walk-through to avoid individual-related bias. Each group included at least one bird expert and one railway expert. For safety reasons, surveys were conducted by following all the safety procedures with the presence of a safety officer from the railway company. Surveys were cancelled when the weather condition was unfavorable, such as heavy rainfalls and poor visibility in rainy and foggy days. The survey trips were carried out at the end of March, April, and July in 2016. Each trip contained 3-4 days of field work, and each day we performed one thorough walking investigation of the transect lines starting from 6 a.m. to 9 a.m. to capture roosting crest ibises flying off their perch. One exception was made on 29/03/2016 due to sudden heavy rainfall in the morning. To compensate, we carried out the survey from 16:45 p.m. to 18:55 p.m. on that day when crested ibises were flying back to their resting places. We recorded the observations with the following information: taxonomic identification, number of birds, time, and flight related information including approximate flight height, crossing behavior, and distance to the bridge. The distance to the bridge for each entry was approximated as the horizontal distance from the position at which those birds were firstly observed to the railway bridge. This distance is recorded as ``$<$ 5m", ``6-10m", ``11-25m", ``25-50m", or ``$>$50m". 

Small passerines were not included in the survey, such as white wagtails (\textit{Motacilla alba}), swallows (\textit{Hirundo rustica}), and house sparrows (\textit{Passer domesticus}). However, the Eurasian magpie (\textit{Pica pica}), also known as the common magpie, was particularly recorded due to its relatively large body size in passerines. We identified each bird to as lowest taxonomic level as possible.

Particular attention was paid to the flight related information. We classified the crossing behaviors into three categories, namely ``Type I - Crossing cover the bridge", ``Type II - Crossing under the bridge" and ``Type III - Non-crossing movements", respectively. The catenary has been implemented at a 5m height, making the gap between it and the rails as a bird-train collision risk area~\citep{garcia2017board}. For collision risk evaluation, the flight height of each Type I crossing behavior was estimated with reference to the catenary supporting poles (5m high). Type II and III were considered to be risk-free activities, which include flying across the bridge between piers, foraging, wading under the bridge, flying alongside and away from it and other non-crossing movements such as resting on tree tops.

For data analysis, we applied two ecological measurement indexes. The relative abundance index (RAI) indicates how abundant one species is among all the encountered species. The encounter rate of a particular species (ER) represents how likely a species would be encountered per unit distance. They are mathematically expressed as:

\begin{equation}
RAI_{i} (\%)= \frac{N_{i}}{T}\times 100
\label{eq1}
\end{equation}

\begin{equation}
ER_{i} (\%)= \frac{N_{i}}{L}\times 100
\label{eq2}
\end{equation}
where $N_{i}$ is the number of observations recorded for the species $i$, $T$ is the total number of identification of all the encountered species, and $L_{j}$ is the total distance covered in the transect. We also applied linear regression modeling techniques in cumulative encounter analysis and $R-Squared$ to verify the goodness-of-fit of the regression models. All data analysis were carried out with MATLAB R2015a.  \color{black}

\section{Results}
\subsection{Estimation of species abundance}
\begin{figure}[h!]
\centering
\includegraphics[width=1\textwidth]{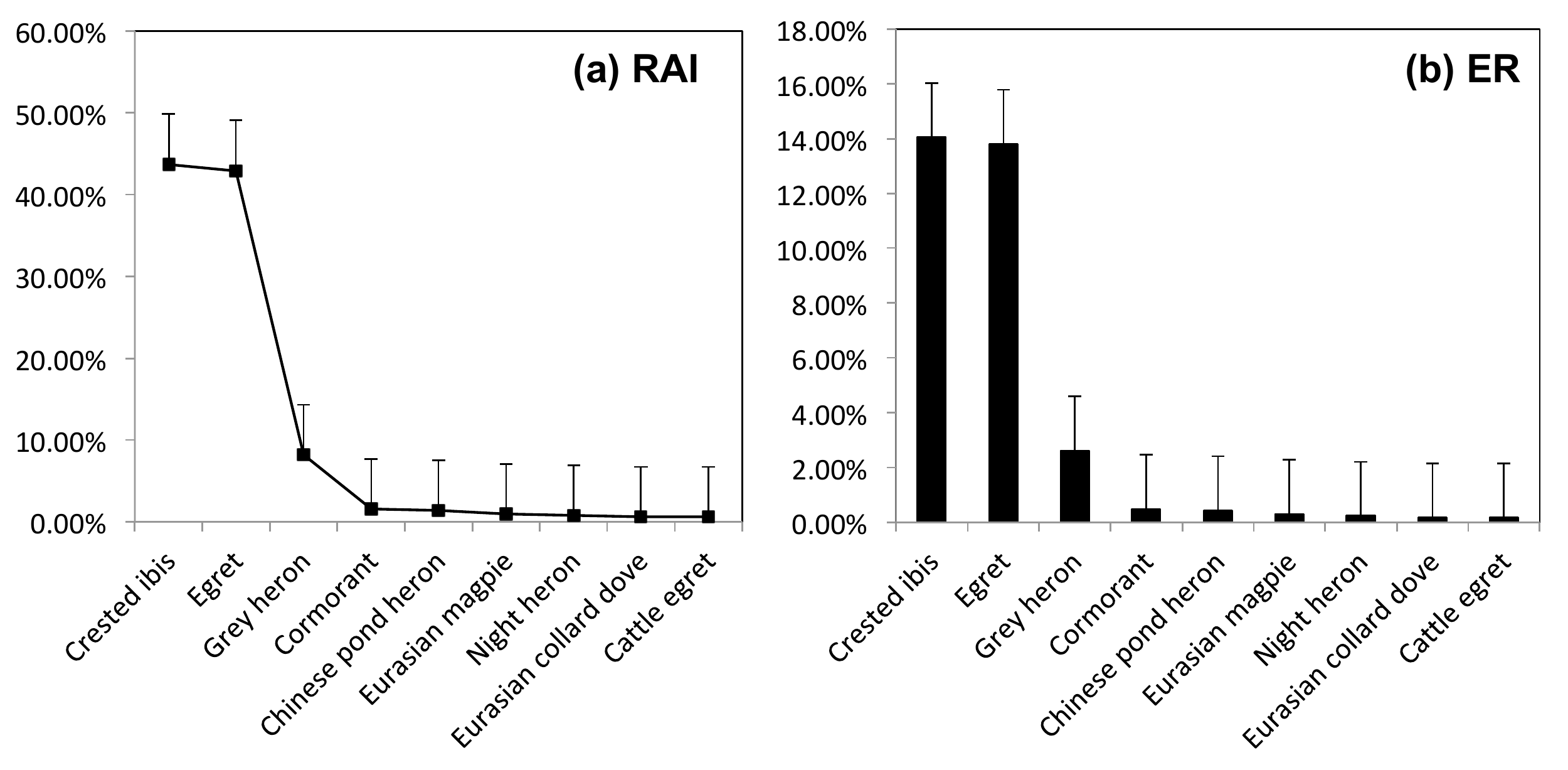}
\caption{\small Encounter Rate (ER) of observed species based on survey data. (color is NOT needed)}
\label{Fig4}       
\end{figure}

Fig.~\ref{Fig4} shows the results of RAI and ER for all the nine medium-sized or large bird species observed during the surveys. Most of them (except common magpie and Eurasian collared dove) are waterfowls and waders. These aquatic birds contribute to a high percentage of the total observations (98\%). As can be seen, the two most observed bird species in the study area were crested ibis (with RAI of 43.69\%) and egret (42.91\%), while other species have much lower RAI values (grey heron, 8.16\%; Cormorant, 1.55\%; Chinese pond heron, 1.36\%). The RAI values for magpie, night heron, Eurasian collared dove, and cattle egret were all less than 1\% (we only observed no more than five of these species). In subplot (b), the ERs of these bird species have a similar trend as found in RAIs. The crested ibis and egret have the top two ER values (crested ibis, 14.06\%; egret, 13.81\%). The rest all have the ER values lower than 3\%.


\begin{figure}[htp!]
\centering
\includegraphics[width=0.65\textwidth]{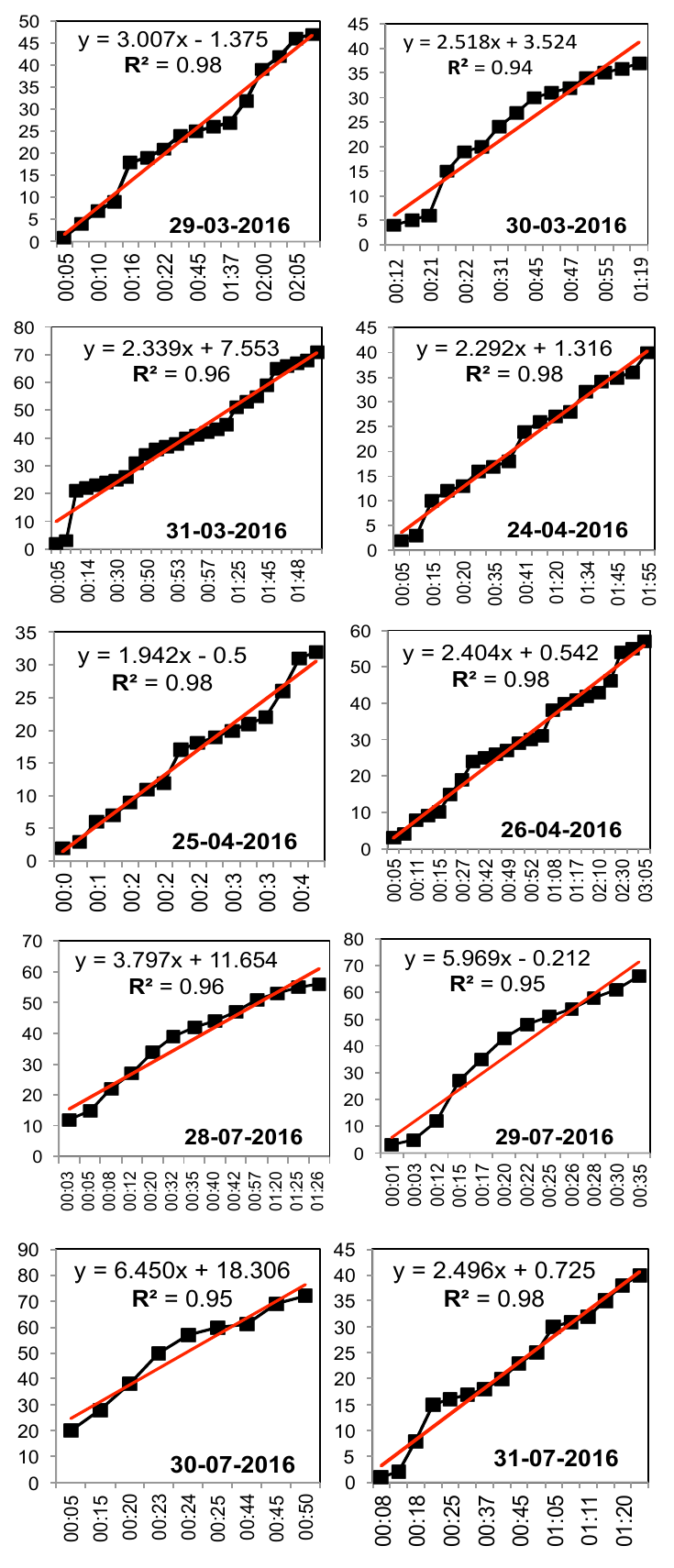}
\caption{\small Cumulative curve of bird observations during the surveys. For each plot, X-axis is Time; Y-axis is Cumulative frequency; Red line is linear regression model; R$^2$ is the adopted measure for goodness of fit.}
\label{Fig6}       
\end{figure}

The cumulative curve of bird observations during each survey is shown in Fig.~\ref{Fig6}. In this analysis, we calculated the cumulative bird observations against the survey duration to investigate the temporal patterns of the species. The results show that all the curves could be approximated by linear regression models with an average $R^2$ of 0.966, which demonstrates that the overall encounter in each survey follows a uniform distribution over time. The mean slope of the regression models ($\bar m=$ 3.321) shows that on average, there were three birds spotted in every five minutes during the surveys. In addition, most of the first detections took place within the first five minutes after the commencement of the survey, except for the results on 30 March 2016 (12 minutes) and 31 July 2016 (eight minutes). 

\begin{figure}[htp!]
\centering
\includegraphics[width=1\textwidth]{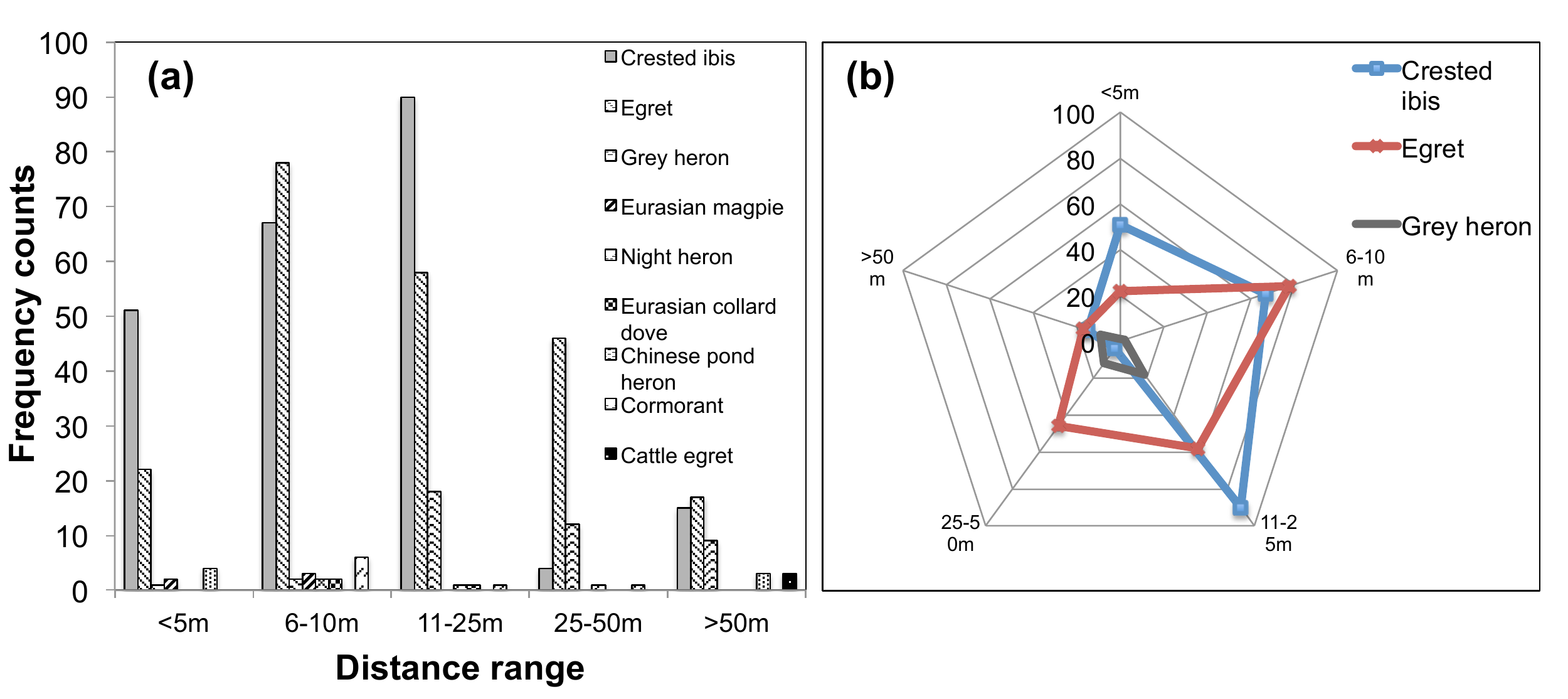}
\caption{\small Analysis of the horizontal distance to the bridge at which those birds were firstly detected. (a) Frequency counts for each bird species with respect to the distance ranges. (b) Spatial distribution of the distance to the bridge.}
\label{Fig8}       
\end{figure}

Fig.~\ref{Fig8} (a) shows the species-level total frequency counts with respect to the distance to the bridge. Most crested ibises (90 observations) were first spotted in the distance range between 11m to 25m. In contrast, most egrets were first observed closer to the bridge, at a distance of 6-10m (78 observations). To take a closer look at the overall distance distribution of the top three species (crested ibis, egret, and grey heron), we provide Fig.~\ref{Fig8} (b). We can see the overall distribution of crested ibises was more skewed and closer to the bridge than that of egrets and grey herons. The majority of crested ibises (91.63\%) were first observed within 25m to the bridge, while the majority of egrets (90.05\%) were first spotted at a more distant location ranging from 6m to 50m. The small magnitude of polygon area for grey herons is due to its few total observations compared to the other two.

\subsection{Bridge crossing}
\begin{figure}[htp!]
\centering
\includegraphics[width=1\textwidth]{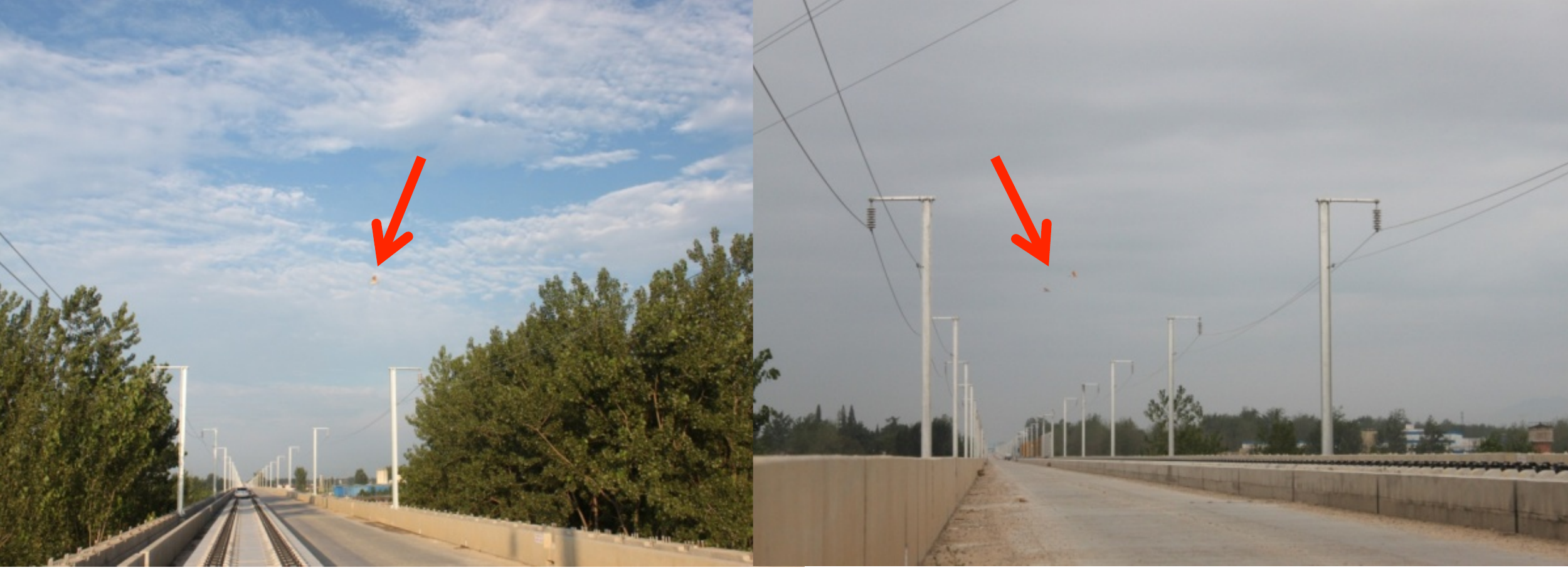}
\caption{\small Type I crossings of crested ibises observed during the surveys. Red arrow shows bird position when flying over the bridge.}
\label{Fig5}       
\end{figure}

\begin{figure}[htp!]
\centering
\includegraphics[width=0.7\textwidth]{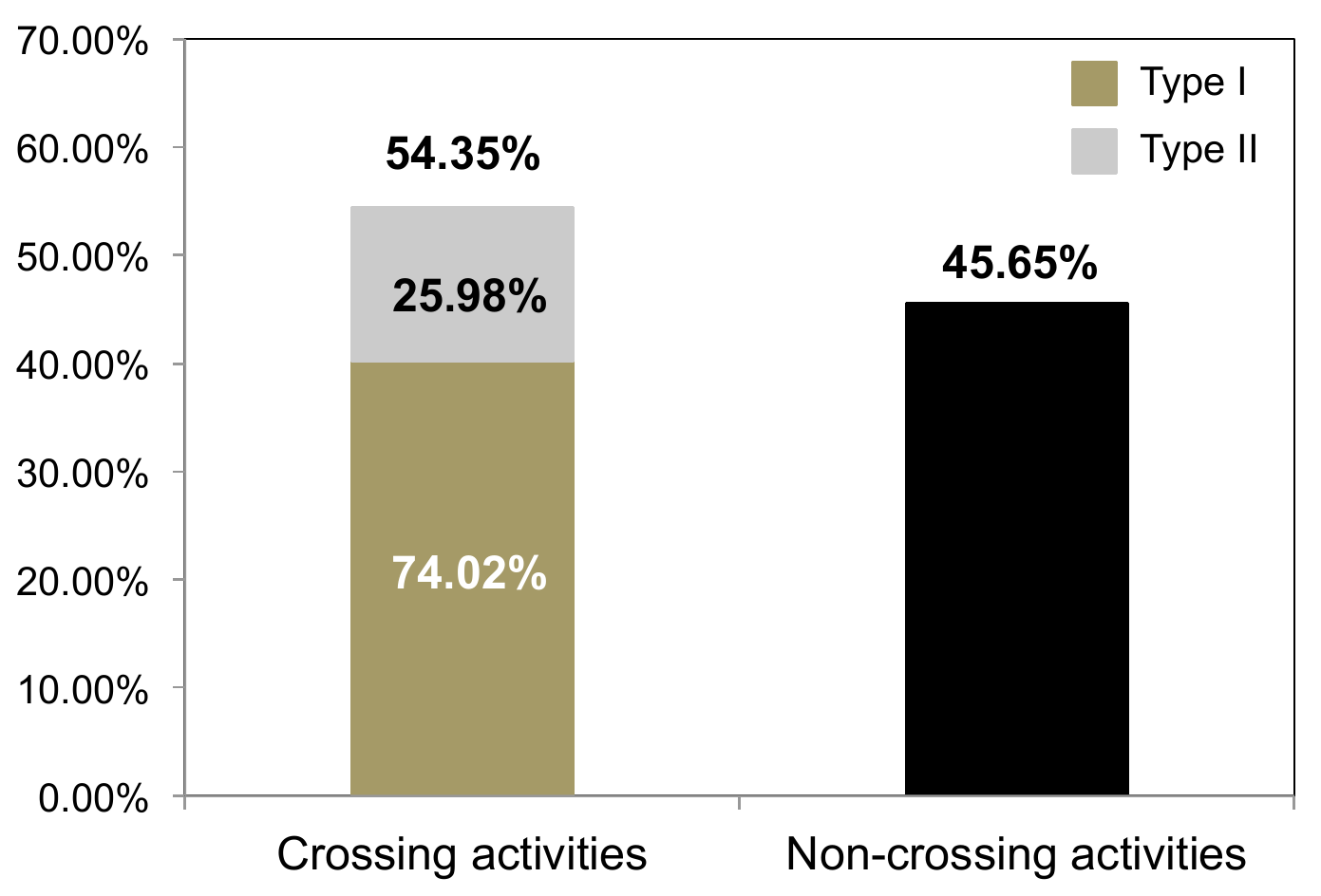}
\caption{\small Crossing and non-crossing activities for all observed bird species}
\label{Fig7}       
\end{figure}

Bird collision risk can be evaluated by analyzing the birds' flight behaviors around the engineering structure (Fig.~\ref{Fig5}). Fig~\ref{Fig7} presents the results on the total crossing (Type I and II) and non-crossing (Type III) behaviors from all observations. Compared to Type III non-crossing behaviors (45.65\%), the demand for crossing the bridge was relatively higher (54.35\%). Among all crossing behaviors, Type I was the dominant type, leading to a ratio of about 7:3 between Type I (74.02\%) and Type II (25.98\%) crossing activities.

Table~\ref{tab4} summarizes these three types of flying behaviors and the average flying heights of each bird species. The non-crossing and crossing activities of crested ibises were equal, but most of the crossings were made over the bridge (100 observations, 89.29\% of total crossing activities of this species) with only 12 crossings were made under the bridge (10.71\% of total crossing activities of this species). Also, this species contributed most to the Type I crossing activities among all species (100 observations, 48.08\% of total 208 Type I crossings). Egrets have the highest number of crossings (137 out of 221). While Type I crossing was also the dominant type for egrets (93 observations, 67.88\% of total crossings of this species and 42.08\% of total 221 observations), the number of Type II crossing of this species was much higher than those of other species (44 observations, 60.27\% in total 73 Type II crossings among all species). From statistics, we found that: (1) Most of the crossing behaviors were made by egrets and crested ibis; (2) Most of the observed birds prefer Type I crossing; (3) Crested ibises have the highest Type I crossing count; and (4) Egrets have the highest Type II crossing count.


Additionally, the average flight heights for each bird species were also recorded in Table~\ref{tab4}. Most of the observed birds flew with an average height above 15 m, except magpies and night herons. Particularly, grey herons and crested ibises were the two species with a much greater flying height, approximately 25.6 m and 20.3 m respectively, which are much higher than the average height of the railway bridge (about 10 m). In contrast, the flight height of egrets was about 16.9 m on average, which is only slightly above the bird-train collision risk area.

\begin{table}[hpt!]
\caption{\small Statistical summary of three types of behaviors at species level}
\label{tab4}
\resizebox{\textwidth}{!}{\begin{tabular}{|l|l|l|l|l|l|l|}
\hline
Species & Average height  & Type III & Type I & Type II & Total & \textbf{Total} \\
&(visually approximated&Non-crossing &crossing&crossing&crossing&\textbf{observation}\\
&during observation)& behavior&&&(Type I+II)&\\ \hline
Crested ibis & 20.3 & 112 & 100 & 12 & 112 & 224\\ \hline
Egret & 16.9 & 84 & 93 & 44 & 137 & 221\\ \hline
Grey heron & 25.6 & 25 & 6 & 11 & 17 & 42\\ \hline
Eurasian magpie & 10.0 & 2 & 2 & 1 & 3 & 5\\ \hline
Night heron & 3.0 & 4 & 0 & 0 & 0 & 4\\ \hline
Eurasian collard dove & 20.0 & 2 & 1 & 0 & 1 & 3\\ \hline
Chinese pond heron & 20.0 & 3 & 2 & 2 & 4 & 7\\ \hline
Cormorant & 17.0 & 1 & 4 & 3 & 7 & 8\\ \hline
Cattle egret & 20.0 & 3 & 0 & 0 & 0 & 3\\ \hline
\textbf{Total count} & - & 236 & 208 & 73 & 281 & 517\\ \hline
\end{tabular}}
\end{table}

\section{Discussion}
A significant amount of research on crested ibis has been contributed from the Far East countries, such as China and Japan, covering a number of habitat- and behavior-related topics, including habitat evaluation and protection~\citep{li2001using,li2002habitat,li2009crested}, reproductive success~\citep{yu2006reproductive,li2018survival}, breeding and nesting~\citep{yu2015breeding,dong2018effects}, and reintroduction programme~\citep{wang2017sustainability,wu2017breeding,wajiki2018estimation}. These studies show that the protection on the crested ibis has been successful and also indicate the need for continuous protection on this bird species.

Thus, our study contributes to the protection of crested ibis from railway ecology perspective. The results indicate that the crested ibis and egret are the two most abundant bird species in the study area. This result confirms the previous investigations that the population of this endangered bird species increased significantly during last decades~\citep{yu2006reproductive}. Also, it has been found that the egret is one of the co-habitants of the crested ibis in this study area. The population of this species has also been thriving in years~\citep{li2014genomic}, and it is one of the common waders found in the Qinling Mountains according to local wildlife census. This estimation result supports the hypothesis that the crested ibis and the egret would relatively be prone to the collision risk in terms of their abundance. Based on the temporal cumulative curves, the resultant linear models and high $R^2$ values imply a strong uniformly-distributed characteristic of birds over time in the study area. The results of average slope ($\bar m=$ 3.321) and the first encounter time ($<$ 5 mins) are in line with the high encounter rate perceived by our surveyors as flocks of birds were frequently spotted. The number of other species from Ardeidae family was low during the surveys. Thus, inferences related to them should be considered with care. We think that further investigations are needed to expound their relative scarcity in the area.

The distance to the roads is recognized as one of the major factors affecting the bird mortality along roads~\citep{piao2016preliminary}. Even though previous studies have shown that deliberate human-induced risks can affect the population growth and range expansion of crested ibises~\citep{sun2016predicting}, they are found to be able to live rather close to human settlements~\citep{hu2016size}. The tolerance capability of crested ibises to human disturbance have been increasing since last decade~\citep{li2002habitat}. The result obtained in the distance analysis supports such claim as we found they were more overall closer to the railway bridge than other bird species. However, a recent theoretical study~\citep{zhang2017influence} has shown that crested ibises prefer to nest in the areas distant from man-made structures, such as high-grade highways, which seems to contradict with the observations in our railway case. One of the possible explanations is that the railway bridge had been completed for quite a while and the train services were not in operation when we conducted the surveys. It is possible that such closeness to the bridge could be due to the fact that these birds have already been used to the existence of the railway bridge. More explanatory studies are certainly needed with the aid of future monitoring work.

The analysis of crossing behaviors is also vital for evaluating bird collision risk. We found that bridge-crossing behaviors accounted for a large portion of the birds' total daily activities in this area. For the crested ibis, while their frequency of crossings was much higher compared to non-crossing behaviors, we still believe the risk for this species to collide with trains could be low. Although most of observed crossings were above the bridge (i.e., Type I ), which may be prone to collision risk, the flight heights of these crested ibises were quite sufficient for safe crossings. This inference can be supported by similar studies in other waterfowl species, such as \citet{godinho2017bird} pointed out that large aquatic birds may be less exposed to collisions because they normally fly higher than others.

However, compared to crested ibises, other bird species might be under higher collision risk according to our findings. Egrets leave much squeezed space above the collision risk area for their crossings with relatively lower flight heights observed in the surveys. To ensure a safe flying-over, mitigation measures could be considered. One possible mitigation measures could be providing obstacles, such as fences. These structures could help to deflect birds' flying trajectory by forcing them to pull up. This is based on the observations that individual birds and bird flocks change flight paths in response to the presence of artificial structures~\citep{zuberogoitia2015testing}. By doing so, larger clearance of the space for crossings would significantly promote the safety associated with. Previous studies have also shown that passerines are one of the major victims in the bird-train collisions~\citep{garcia2017board,malo2017cross}. Therefore, when considering the interests of multiple species, while our results together with similar previous studies suggest that the bird collision risk associated with railways could be relatively low compared to other sources of anthropogenic mortality~\citep{loss2015direct}, we still recommend that barrier-like structures, such as fences between catenary and rails on both sides of the XCHR should be considered at particular sections. 

We also notice that the ratio of total Type I and II crossings obtained in our results is about 7:3, which is different from the value obtained in a similar previous study.~\citet{godinho2017bird} found this ratio is about 9:1 in their study case. We think bridge crossing behaviors might vary among cases. Different bird species, landscapes, number of observations, seasons, and railway bridges could be potential influencing factors. Thus, we agree with the claim that reaching a generalization on birds' bridge-crossing studies could be very difficult~\citep{godinho2017bird}.

From the technical perspective, the methods applied in this study are all field-based and relatively low-cost yet labor-intensive. There are other equipments and technologies that are potentially helpful to enhance this study and facilitate the future monitoring work. One is to utilize information technology and automation systems. For example, on-board cameras could be implemented for monitoring the in-situ situations of bird-train collisions in the future. This method can also provide useful information such as escaping distance and avoidance behaviors~\citep{garcia2017board}. Also, infra-red cameras could also be considered at fixed points and it is specially useful for long-term and night-time monitoring~\citep{zhang2019field}. Artificial intelligent systems could also help to extract valuable information from massive digitalized recordings and images so that labor costs could be significantly reduced~\citep{longmore2017adapting}. However, these high-tech methods have their own disadvantages such as high cost. One practical strategy is to effectively combine those smart devices with field trips. From this perspective, we see huge potential for improving the investigation methods in railway ecology studies.

Some limitations of this study can be identified. Firstly, even though most of the observed birds are diurnal, the timing of the surveys may introduce hidden bias to other bird species. For example, the estimation results may alter during daytime and nighttime as some bird species are night flyers (e.g., owls and nightjars). The field surveys were conducted in summer and autumn due to scheduling constraints of the project. This limitation may have omitted the effects of seasonal dynamics on migratory birds~\citep{mata2009seasonal,serronha2013towards}. Secondly, the surveys were carried out at monthly intervals and only three to four days per month, which may influence abundance estimations due to undersampling~\citep{santos2015sampling}. Finally, surveys during heavy rainy and foggy days were avoided in this study. However, humid weather might trigger unusual flying patterns~\citep{boyle2010storms,kirsch2015observation}, which is not considered in this study. Therefore, our conclusion about birds' bridge-crossing behaviors should be accepted with limited generality. To mitigate the negative effects of these limitations on the results, we recommend future work could consider covering a longer time span for field observations with respect to a wider range of multiple species.


\section{Conclusion}
In this study, we report a recent bird survey project associated with the Xi'an-Chengdu high-speed railway (XCHR) to demonstrate how we evaluate the bird collision risk of the crested ibis and local birds. The study provides a pioneering investigation on the potential conflicts between crested ibises and high-speed trains and highlights some key aspects in the study of bird crossing behavior associated with railway bridges that may lead to anthropogenic mortality. Three findings are summarized as follows:
\begin{itemize}
\item The crested ibis and the egret are reported to be the two most abundant observed bird species in the study area. The RAI values of these two species are 43.69\% and 42.91\%, and the their ER values are 14.06\% and 13.81\%, respectively.

\item Crested ibises were largely first spotted at a closer distance to the railway bridge as 91.63\% of them were detected within 25m of the vicinity of the bridge.

\item The ratio between total Type I and II flying behaviors was about 7 : 3. Crested ibises were found to prefer crossing from above (89.29\% of their total crossing movements). In contrast, more crossing activities were contributed by egrets (137 observations) and they are responsible for 60.27\% of all Type II observations. 
\end{itemize}

Based on the findings and analysis, we believe that the possibility of crested ibises to collide with bullet trains could be relatively low. Nevertheless, we recommend the implementation of fences on both sides of the railway bridge for a greater safety margin to multiple bird species. Future work will focus on identifying the sections at which fences need to be implemented for XCHR to prevent possible bird-train collision risk.


\newpage
\singlespacing
\section*{Fundings}
This research was supported by the Major Scientific Research Project of Shaanxi Academy of Science [grant number: 2016K-04] and Future Resilient Systems at the Singapore-ETH Centre, which was established collaboratively between ETH Zurich and Singapore's National Research Foundation (grant number: FI 370074011) under its Campus for Research Excellence and Technological Enterprise programme.

\section*{Author contributions and statement of competing interests}
All of the authors contributed to the conception and design of this study, and have read and approved the final manuscript. The authors declare no conflicts of interest.

\section*{Acknowledgements}
The authors would like to express their deepest gratitude to all of the volunteers and coordinators who participated in and contributed to the field investigations.

\newpage


\cleardoublepage

\end{document}